\documentstyle[12pt]{article}
\def\d{\delta}

\def\L{\Lambda}

\newcommand{\p}[1]{(\ref{#1})}
\begin{document}
\thispagestyle{empty}
\renewcommand{\thefootnote}{\fnsymbol{footnote}}
\begin{flushright}
Preprint DFPD 97/TH/31\\
hep-th/9707044\\
July 1997
\end{flushright}

\vspace{3truecm}
\begin{center}
{\large\bf Covariant Actions for the Bosonic Sector of D=10 IIB
Supergravity}

\vspace{1cm}
Gianguido Dall'Agata$^1$, Kurt
Lechner$^1$\footnote{kurt.lechner@pd.infn.it} and Dmitri
Sorokin$^2$\footnote{dsorokin@kipt.kharkov.ua}

\vspace{0.5cm}
${}^1${\it Universit\`a Degli Studi Di Padova,
Dipartimento Di Fisica ``Galileo Galilei''\\
ed INFN, Sezione Di Padova,
Via F. Marzolo, 8, 35131 Padova, Italia}

\vspace{0.5cm}
${}^2${\it National Science Center,
Kharkov Institute of Physics and Technology,\\
Kharkov, 310108, Ukraine}

\vspace{1.cm}
{\bf Abstract}
\end{center}
Covariant actions for the bosonic fields of D=10 IIB supergravity
are constructed with the help of a single auxiliary scalar field and
in a formulation with an infinite series of auxiliary
(anti)--self--dual 5-form fields.

\bigskip
\bigskip
\renewcommand{\thefootnote}{\arabic{footnote}}
\setcounter{footnote}0

The construction of a complete action for chiral $N=2$ (or IIB)
supergravity
in ten--dimensional space--time \cite{IIB} is a long standing problem,
which has become topical again in connection with a ``duality
revolution".
Its solution has been hampered by the presence in the bosonic sector of
the theory of a four--form gauge field $A^{(4)}(x)$ whose five--form
field
strength $M= dA^{(4)}+...$ is self--dual, i.e. $M=M^*$ (where * denotes
the Hodge conjugation of the differential forms). This field comes from
the Ramond--Ramond (RR) sector of IIB superstring theory. The problem is
to find a $D=10$ covariant action from which the self--duality condition
would follow as an $A^{(4)}$ field equation of motion.

An action proposed for IIB supergravity in \cite{km} is based on a
modification of the Siegel approach \cite{siegel} to a space--time
covariant
Lagrangian description of self--dual fields. An unsatisfactory feature
of
this construction is that Lagrange multiplier fields which are used to
get
the self--duality condition cannot be completely eliminated by local
symmetries and equations of motion of the model, and the treatment
of
remaining auxiliary degrees of freedom is not clear.

Another possibility of solving the
problem of the self--dual field is to use a covariant McClain--Wu--Yu
approach which involves infinite number of auxiliary fields
\cite{inf,ber1,bk}.
An interesting observation concerning this construction is that it
arises as an effective action of a closed superstring field theory
\cite{ber1}. Because of an infinite series of fields one may encounter
technical complications when generalizing such a model to a locally
supersymmetric case\footnote{Note, however, that in \cite{ber2} it has
been
shown how to construct in this approach superfield actions for $D=4$
supersymmetric Maxwell theory with manifest dualities.}.

In this note we propose a $D=10$ covariant action for the bosonic sector
of IIB supergravity, where to get manifest space--time covariance of the
self--dual field part of the action only one scalar auxiliary field is
used. Such a covariant approach to deal with self--dual fields was
proposed in \cite{pst1,pst2} and has been applied to the
duality--symmetric
description of (super)Maxwell theory \cite{pst1}
coupled to electric and magnetic
sources \cite{source}, and to the $SL(2,R)\times SO(6,22)$ invariant
effective field theory action in
$D=4$ of a toroidally compactified heterotic string \cite{pst1}. This
approach has proved to be pretty helpful in constructing a complete
action for the M--theory 5--brane propagating in a $D=11$
supergravity background \cite{m5} and considering its K3
compactification
\cite{k3}. A relationship between the single scalar field approach and
the
McClain--Wu--Yu approach has been established in \cite{pst2}, which
allows one
to straightforwardly get a IIB supergravity action in the latter
formulation. We present this action in the conclusion of the article.

The covariant action for the bosonic sector of $D=10$ IIB supergravity
which one gets with the use of a single scalar auxiliary field
has the following form
$$
S=\int d^{10}x\sqrt{-g}[R-
2\partial_m\phi\partial^m\phi-2e^{2\phi}
\partial_m\phi'\partial^m\phi'-{1\over 3}e^{-\phi}H_{lmn}H^{lmn}
$$
\begin{equation}\label{1}
-{1\over 3}(H'_{lmn}-\phi'H_{lmn})(H'^{~lmn}-\phi'H^{~lmn})]
\end{equation}
$$
-{1\over 6}\int d^{10}x{{\sqrt{-g}}\over{\partial_ra\partial^ra}}
\partial^la(x)M^*_{lm_1...m_4}{\cal M}^{m_1...m_4p}\partial_pa(x)
-4\int A^{(4)}\wedge H\wedge H',
$$
where $g_{mn}(x)$ (m,n=0,...,9) is a $D=10$ space--time metric, $R(x)$
is
a $D=10$ scalar curvature, $\phi(x)$ and $\phi'(x)$ are,
respectively, the Neveu--Schwarz (NS) dilaton and the RR scalar;
$H=dB$ and $H'=dB'$ are the three--form field strengths of,
respectively,
the NS two--form gauge field $B(x)$ and the RR
two--from gauge field $B'(x)$,
\begin{equation}\label{M}
M=dA^{(4)}+{1\over 2}B\wedge H'-
{1\over 2}B'\wedge H
\end{equation}
is the five--form
field strength of the RR gauge field $A^{(4)}(x)$ extended with the
Chern--Simons--like term constructed of the two--form fields, and ${\cal
M}=M-M^*$ is the anti--self--dual part of $M(x)$. Finally $a(x)$ is the
auxiliary scalar field ensuring the manifest $D=10$ covariance of the
$A^{(4)}$--part of the action, which would otherwise be lost
(see \cite{ss} and references therein).
Entering the action in a nonpolynomial way the field $a(x)$ points to
a
nontrivial topological structure of the self--dual field
configurations.
The form of the $M(x)$ terms as written
in \p{1} is convenient for making the symmetry analysis of the
action and getting the equations of motion, though, one can also present
the $M(x)$ part of the action \p{1} in more conventional form
\begin{equation}\label{2}
S_M=-\int d^{10}x\sqrt{-g}\big[{1\over{60}}
M_{m_1...m_5}M^{m_1...m_5}-{1\over{12\partial_ra\partial^ra}}
\partial^la(x){\cal M}_{lm_1...m_4}{\cal
M}^{m_1...m_4p}\partial_pa(x)\big]
\end{equation}
$$
-4\int A^{(4)}\wedge H\wedge H'
$$
which is the sum of the kinetic term of $A^{(4)}$, the term quadratic
in the anti--self--dual tensor ${\cal M}(x)$ and the Chern--Simons term.
In
this second form \p{2} the action \p{1} is the same
as the one written in \cite{t} except that it contains the second term
in
\p{2}.
This difference is crucial since, while in \cite{t} the self--duality
condition
\begin{equation}\label{3}
{\cal M}=M~-~M^*=0
\end{equation}
had to be imposed into the theory ``by hand", with this modification
Eq. \p{3} becomes the consequence of the $A^{(4)}$ equations of motion.
This has been demonstrated in detail for various models containing
self--dual fields \cite{pst1}--\cite{ss} and we address the reader to
these
references. Let us only note that, in addition to $D=10$ general
coordinate
transformations, standard gauge transformations of $B$, $B'$ and
$A^{(4)}$ and global $SL(2,{\rm R})$ duality mixing of $\phi$ and
$\phi'$ and of $B$ and $B'$  \cite{IIB,t},
the action \p{1} is also invariant under
the following local transformations of $A^{(4)}$ and $a(x)$:
\begin{equation}\label{4}
\d A^{(4)}=da\wedge\varphi^{(3)}, \qquad \delta a(x)=0;
\end{equation}
\begin{equation}\label{5}
\d a(x)=\varphi(x), \qquad \delta
A^{(4)}_{mnpq}={\varphi\over{(\partial a)^2}}
{\cal M}_{mnpqr}\partial^ra.
\end{equation}
where $\varphi^{(3)}(x)$ is a three--form and $\varphi(x)$ is a scalar
gauge
parameter.

The local symmetry of the action under the transformations \p{4}
reflects the
fact that $A^{(4)}$ is self--dual on the mass shell, i.e. it contains
twice
less (namely 35) physical degrees of freedom in comparison with an
ordinary
four--form abelian field in $D=10$.

The transformations \p{5} allow one to gauge fix $a(x)$ at the expense
of the manifest $D=10$ covariance \cite{pst1,pst2} of the action \p{1}.
Note that the gauge fixing of \p{5} where $\partial_ma$ would be
lightlike
(i.e.  $(\partial a)^2=0$) is inadmissible because of the presence of
the
norm of the vector $\partial_ma$ in the denominator of the action \p{1}
or
\p{2}.  Thus globally $da(x)$ is a closed form but not exact.

An interesting feature of the symmetries \p{4} and \p{5} is that, as in
the  case of the M--theory five--brane \cite{m5}, they fix the relative
coefficients of the term containing the auxiliary field $a(x)$ and of
the Chern--Simons (the last) term in the action \p{1}.

The equations of motion one gets from the action \p{1} are the same as
the
covariant equations of motion of the bosonic fields of $D=10$ IIB
supergravity which have been known for a long time \cite{IIB}.
The field $a(x)$ disappears from them. As one can directly see from
\p{2}
the $a(x)$ dependent terms in the field equations are always
proportional
to the anti--self--dual tensor ${\cal M}(x)$ which vanishes on the
mass shell (Eq. \p{3}).

For completeness we also present how the self--dual part of the IIB
supergravity action looks like in a simplified covariant formulation
with
an infinite
number of auxiliary 5-form fields $\L^{(n)}_{m_1...m_5}$ $(n=1,...
\infty)$
self--dual for odd $n$ and anti--self--dual for even $n$:
\begin{equation}\label{8}
S_M=-\int d^{10}x\sqrt{-g}[{1\over{60}}
M_{m_1...m_5}M^{m_1...m_5}-\L^{(1)}_{m_1...m_5}{M}^{m_1...m_5}
+\sum_{n=0}^{\infty}(-1)^n\L^{(n+1)}\L^{(n+2)}]
\end{equation}
$$
-4\int A^{(4)}\wedge H\wedge H'.
$$
As it has been shown in detail for free self--dual fields in $D=2p+2$
\cite{inf}--\cite{ber2}, there is an infinite parameter local symmetry
(valid also when $A^{(4)}$ couples to $B$ and $B'$ as in \p{8})
which allows one (at the level of equations of motion) to eliminate all
the
Lagrange multipliers and remain with the self--duality condition \p{3}.

In \cite{pst2} it has been observed that a self--dual action of the form
\p{2} (or \p{1}) can be viewed as a consistent covariant truncation
(or a gauge fixing of the infinite--parameter local symmetry) of \p{8},
where
all $\L^{(n)}$ with $n>1$ are set equal to zero and
$$
\L^{(1)}_{m_1...m_5}={1\over{6\partial_ra\partial^ra}}\big(
\partial^la(x){\cal M}_{l[m_1...m_4}\partial_{m_5]}a(x)
+{1\over{4!}}\epsilon_{m_1...m_5m_6...m_{10}}
\partial_la(x){\cal M}^{lm_6...m_9}\partial^{m_{10}}a(x)\big).
$$

It should be possible to promote the action \p{1} to a complete IIB
supergravity action, where  off--shell supersymmetric transformations of
fermionic fields will include nonconventional terms proportional to the
anti--self--dual tensor ${\cal M}$, which vanish when Eq. \p{3} is taken
into account, as the example of the duality symmetric
super--Maxwell model teaches us \cite{ss,pst1}.

\bigskip
\noindent
{\bf Acknowledgments}. D.S. is thankful to M. Green, B. Julia, P.
Townsend
and B. de Wit for discussion which stimulated the writing of this note.
The authors
would also like to thank N. Berkovits, and M. Tonin for
interest in this work and discussion.
Work of
K.L. was supported by the European Commission TMR programme
ERBFMRX--CT96--0045 to which K.L. is associated.
Work of D.S. was partially supported by research grants of
the Ministry of Science and Technology of Ukraine and the
INTAS Grants N 93--127--ext. and N 93--493--ext.


\begin{thebibliography}{99}
\bibitem{IIB}
Schwarz J. H. 1983 {\sl Nucl. Phys.} {\bf B226} 269;\\
Schwarz J. H. and West P. C. 1983 {\sl Phys. Lett.} {\bf 126B} 301;\\
Howe P. S. and West P. C. 1984 {\sl Nucl. Phys.} {\bf B238} 181.
\bibitem{km}
Kavalov An. R. and Mkrtchyan R. L. 1987 {\sl Soviet J. Nucl. Phys.}
{\bf 46} 1246; 1990 {\sl Nucl. Phys.} {\bf B331} 391.
\bibitem{siegel}
Siegel W. 1984 {\sl Nucl. Phys.} {\bf B234} 307.
\bibitem{inf}
McClain B., Wu Y. S. and Yu F. 1990 {\sl Nucl. Phys.} {\bf B343} 689;\\
Wotzasek C. 1991 {\sl Phys. Rev. Lett.} {\bf 66} 129;\\
Martin I. and Restuccia A. 1994 {\sl Phys. Lett.} {\bf B323} 311;\\
Devecchi F. P. and M. Henneaux M. 1996 {\sl Phys. Rev.} {\bf D45} 1606.
\bibitem{ber1}
Berkovits N. 1996 {\sl Phys. Lett.} {\bf 388B} 743, hep-th/9607070.
\bibitem{bk}
Bengtsson I. and Kleppe A., On chiral p--forms, hep-th/9609102.
\bibitem{ber2}
Berkovits N. 1997 {\sl Phys. Lett.} {\bf 398B} 79.
\bibitem{pst1}
Pasti P., Sorokin D. and Tonin M. 1995 {\sl Phys. Lett.} {\bf
B352} 59;
1995 {\sl Phys. Rev.} {\bf D52} R4277,
hep-th/9506109.
\bibitem{pst2}
Pasti P., Sorokin D. and Tonin M. 1997 {\sl Phys. Rev.} {\bf D55} 6292,
hep-th/9611100.
\bibitem{source}
Medina R. and Berkovits N. Pasti--Sorokin--Tonin Actions in the Presence
of Sources, hep-th/9704093.
\bibitem{m5}
Pasti P., Sorokin D. and Tonin M. 1997 {\sl Phys. Lett.} {\bf 398B} 41,
hep--th/9701037;\\
Bandos I., Lechner K., Nurmagambetov A., Pasti P., Sorokin D. and Tonin
M.
1997 {\sl Phys. Rev. Lett.} {\bf 78} 4332, hep--th/9701149;\\
Aganagic M., Park J.,
Popescu C., and Schwarz J. H., World--Volume Action of the M Theory
Five--Brane, hep--th/9701166.
\bibitem{k3}
Cherkis S. and Schwarz J. H., Wrapping the M theory Five--Brane
on K3, hep--th/9703062.
\bibitem{ss}
Schwarz J. H. and Sen A. 1994 {\sl Nucl. Phys.} {\bf B411} 33.
\bibitem{t}
Townsend P. K., Four Lectures on M--theory, hep--th/9612121.
\end{thebibliography}
\end{document}